\documentclass[runningheads,a4paper]{llncs}
\usepackage{fancyvrb}
\usepackage{fixltx2e}
\usepackage{amssymb}
\usepackage{graphicx}
\usepackage{listings}
\usepackage{multirow}
\usepackage{url}

\mathchardef\UrlBreakPenalty=10

\usepackage{fancyvrb,hyperref}
\newcommand{\URL}[1]{\url{\detokenize{#1}}}

\usepackage{cite}
\urldef{\mailsa}\path|{hany, mln}@cs.odu.edu|    
\newcommand{\keywords}[1]{\par\addvspace\baselineskip
\noindent\keywordname\enspace\ignorespaces#1}
\usepackage{colortbl}
\usepackage{alltt}

\begin{document}

\clubpenalty=10000 
\widowpenalty = 10000

\mainmatter

\title{Resurrecting My Revolution}
\subtitle{Using Social Link Neighborhood in Bringing\\ Context to the Disappearing Web}

\author{Hany M. SalahEldeen\and Michael L. Nelson}

\institute{Old Dominion University, Department of Computer Science\\
Norfolk VA, 23529, USA\\
\mailsa\\}

\maketitle

\begin{abstract}
In previous work we reported that resources linked in tweets 
disappeared at the rate of 11\% in the first year followed by 7.3\% each 
year afterwards.  We also found that in the first year 6.7\%, and 14.6\% in each subsequent year, of the resources were archived in public web archives. In 
this paper we revisit the same dataset of tweets and find that our prior model still holds and the calculated error for estimating percentages missing was about 4\%, but we found 
the rate of archiving produced a higher error of about 11.5\%.  We also discovered that resources have disappeared from the 
archives themselves (7.89\%) as well as reappeared on the live web after being declared missing (6.54\%). 
We have also tested the availability of the tweets themselves and found that 10.34\% have disappeared from the live web.  To mitigate the loss 
of resources on the live web, we propose the use of a ``tweet 
signature''.  Using the Topsy API, we extract the top five most frequent 
terms from the union of all tweets about a resource, and use these five 
terms as a query to Google. We found that using tweet signatures results in discovering replacement resources with 70+\% textual similarity to the missing resource 41\% of 
the time.

\end{abstract}

\section{Introduction}
Microblogging services like Twitter have evolved from merely posting a status or quote to an intra-user interaction tool that connect 
celebrities, politicians, and others to the public. They have also evolved to act as a narration tool and an information exchange describing current publicly recognized events 
and incidents. In 2011, during the Egyptian revolution, thousands of posts and resources were shared during the 18 days of the uprising. These resources could have crucial 
value in narrating the personal experience during this historic event, acting as a first draft of history written by the public. 

In our previous work, we proved that shared resources on the web are prone to loss and disappearance at nearly constant rate \cite{tpdl2012}. We found that after only one year we lost nearly 11\% of the 
resources linked in social posts and continued to lose an average of 7.3\% yearly. In some cases, this 
disappearance is not catastrophic as we can rely on the public archives to retrieve a snapshot of the resource to fill into the place of the missing resource. In another 
study we measured how much of the web is archived and found that 16\%--79\% of URIs have at least one archived copy \cite{Ainsworth:2011:MWA:1998076.1998100}. Unfortunately, there is still a large percentage of the web that is not archived and 
thus a huge amount of resources are not archived and prone to total loss upon disappearance from the live web. 

This evolution in the role of social media and the ease of reader interaction and dissemination could be used as a possible solution to metigate or prevent the loss of the unarchived 
shared resources. Fortunately, when a user tweets or shares a link, it leaves behind a trail of copies, links, likes, comments, other shares. 
If the shared resource is later gone these traces, in most cases, still persist. Thus, in this paper we investigate if the other tweets that also linked to 
the resource can be mined to provide enough context to discover similar resources that can be used as a substitute for the missing resource. To 
do this, in this study we extract up to the 500 most recent tweets about linked URIs and we propose a method of finding the social link neighborhood of the resource we are attempting to reconstruct. This link neighborhood 
could be mined for identifiers and alternative related resources. 

\section{Related Work}
Social media has been the focus of numerous studies in the last decade. Twitter, for example, was analyzed by Kwak et al. where they aimed to identify the characteristics of the Twittersphere, 
retweeting, and the diffusion speed of posts by using algorithms like PageRank in ranking users \cite{kwak}. Bakshy et al. investigated 1.6 million users along with the tweet diffusion 
events to identify influencers on Twitter and their effect in content spread \cite{influencers}. To answer questions in regards to the production, flow, and consumption of information on Twitter, Wu et al. 
analyzed the intra-user interactions and found that nearly 20K elite users are responsible for generation of nearly 50\% of URLs shared \cite{wu}. Intuitively, this shows that popularity plays 
an important role in the content disseminated. They also found that type of the content published and the type of users broadcasting this content affect the lifespan of the tweet activity.

Along with understanding the nature of the social media, researchers analyzed user behavior on the social networks in general. By analyzing user activity click logs, Beneventu et al. aimed to get a 
better understanding of social interactions social browsing patterns \cite{patterns}. Zhao and Rosson aimed to explore the reasons of how and why people use Twitter and this use's impact on informal communication at work \cite{zhao}. 
Following the how and the why, Gill et al. attempted to answer the next question of what is the user-generated content is about by investigating personal weblogs to detect the effects of personality, topic type, and the general motivation in published blogs \cite{blogging}. 
Yang and Counts investigated the information diffusion speed, scale and range in Twitter and how they could be predicted \cite{yang}.

This in-depth analysis and study of the social media, its nature, the information dissemination patterns, and the user behavior and interaction paved the way for the researchers to have a better understanding of how the social 
media played a major role in narrating publicly significant events. These studies prove that user-generated content in social media is of crucial importance and can be considered the first draft of history. Vieweg et al. analyzed two natural hazard events (the Oklahoma grass fires and the red River floods in 2009) and how microblogging contributed in raising the situational 
awareness of the public \cite{hazard}. Starbird and Palen analyzed how the crowd interact with politically sensitive context regarding the Egyptian revolution of 2011 \cite{egypt}. Starbird et al. in another study utilized collaborative 
filtering techniques for identifying social media users who are most likely to be on the ground during a mass disruption event \cite{mass}. Mark et al. investigated weblogs to examine societal interactions to a disaster over 
time and how they reflect the collective public view towards this disaster \cite{wardiary}.

In our previous work we showed that this content is vulnerable to loss. Similar to regular 
web content and websites, there are several reasons explaining this loss. McCown et al. analyzed some of the reasons behind the disappearance and reappearance of websites \cite{lost}. McCown and Nelson also examined several techniques to counter 
the loss prior to its occurrence in social networking websites like Facebook \cite{facebook}. As for regular web pages, Klein and Nelson analyzed the means of using lexical signatures to rediscover missing web pages \cite{lexical}. Given that the web 
resource itself might not be available for analysis or might be costly to extract, several studies investigated other alternatives to having the resource itself. Other studies investigated the use of the page's URL to aid web page categorization without 
resorting to the have the webpage itself \cite{kan,Baykan}. Xiaoguang et al. utilized class information from neighboring pages in the link graph to aid the classification \cite{Xiaoguang}.

\section{Existence and Stability of Shared Resources}
We start our analysis by revisiting the experiment conducted in March of 2012, in which we modeled the existence of shared resources on the live web and the public archives. 
In that experiment, we examined six publicly-recognized events that occurred between June 2009 and March 2012, extracting six sets of corresponding social posts. Each of the selected posts include an linked resource and hashtags related to the events. 
Consequently, we tested the existence of the embedded resources on the live web and in the public archives. 
After calculating the percentages lost and archived we estimated the existence as a function of time. In this paper, 
we start by revisiting this year-old estimation model and checking its validity after a year before proceeding with our analysis of reappearance and extracting the social context of 
the missing resources. Then we investigated how this context could be utilized in guiding the search in extracting the best possible replacement for the missing resource.

\subsection{Revisiting Existence}
\begin{figure}
\centering
\includegraphics[scale=0.3]{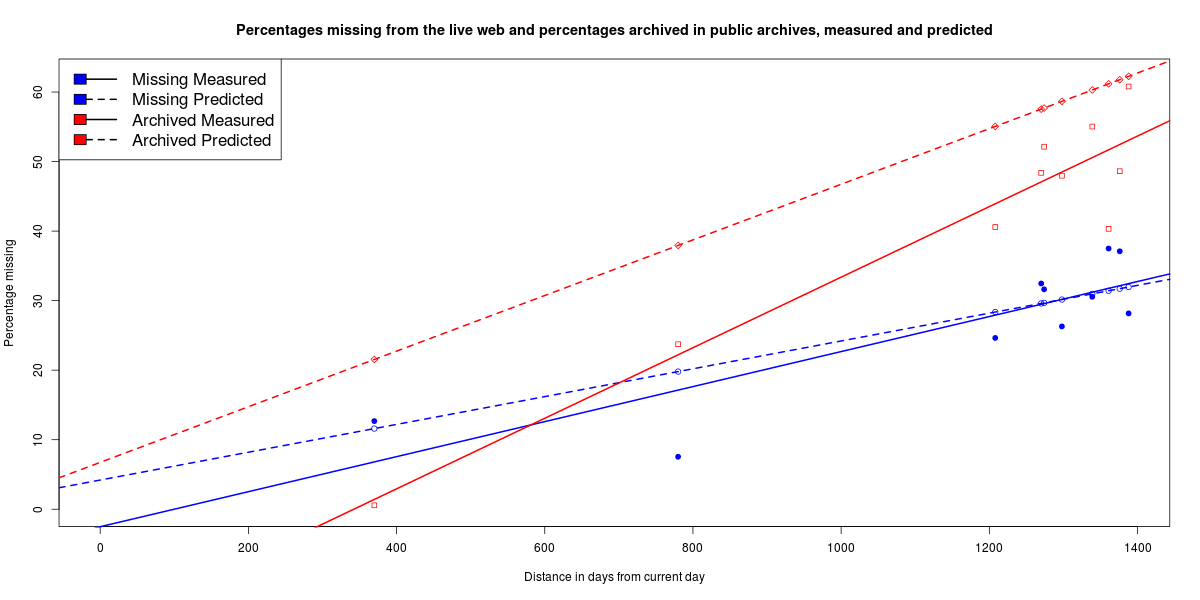}
\caption{Measured and predicted percentages of resources missing and archived for each dataset and the corresponding linear regression.}
\label{fig:afterayear}
\end{figure}

In the 2012 model, we found a nearly linear relationship between the number of resources missing from the web and time (equation \ref{eq:prediction}), and a 
less linear relationship between the amount archived and time (equation \ref{eq:archived}). 
\begin{equation}\label{eq:prediction}
Content\hspace{0.8 mm}Lost\hspace{0.8 mm}Percentage = 0.02 (Age\hspace{0.8 mm}in\hspace{0.8 mm}days) + 4.20
\end{equation}
\begin{equation}\label{eq:archived}
Content\hspace{0.8 mm}Archived\hspace{0.8 mm}Percentage= 0.04 (Age\hspace{0.8 mm}in\hspace{0.8 mm}days) + 6.74
\end{equation}
As a year has passed, we need to analyze our findings and the estimation calculated to see if it still 
matches our prediction. For each of the six datasets investigated, we repeat the same experiment of analyzing the existence of each of the resources on the live web. A 
resource is deemed missing if its HTTP responses terminate in something other than 200, including ``soft 404s'' \cite{baryousef}.
Table \ref{tab:afterayear} shows the results from repeating the experiment, the predicted calculated values based on our model, and the corresponding errors. 
Figure \ref{fig:afterayear} illustrates the measured and the estimated plots for the missing resources. The standard error is 4.15\% which shows that our model matched reality.

\begin{center}
  \begin{table}
      \begin{tabular}[r]{ p{0.25cm} | p{1.2cm}|| p{0.8cm} | p{0.8cm} || p{0.8cm} | p{0.8cm} || p{0.8cm} | p{0.8cm} || p{0.8cm} | p{0.8cm} || p{0.5cm} || p{0.8cm} |}
      \cline{2-12} 
      & &\multicolumn{2}{c||}{\textbf{  MJ  }} & \multicolumn{2}{c||}{\textbf{   Iran   }} & \multicolumn{2}{c||}{\textbf{  H1N1  }} &  \multicolumn{2}{c||}{\textbf{ 
Obama  }} & \multicolumn{1}{c||}{\textbf{Egypt}} & \multicolumn{1}{c|}{\textbf{Syria}}\\ \cline{2-12} \cline{2-12}
      \multirow{7}{*}{\rotatebox{90}{ \scriptsize \textbf{Archived \hskip 7mm Missing} }} & \scriptsize Measured & \tiny 37.10\% & \tiny  37.50\% & \tiny  28.17\%  & \tiny 30.56\% & \tiny 26.29\% & \tiny  31.62\% & \tiny  32.47\%  & \tiny 24.64\% & \tiny  7.55\% & \tiny 12.68\% \\ \cline{2-12}
      & \scriptsize Predicted & \tiny 31.72\% & \tiny  31.42\% & \tiny  31.96\%  & \tiny 30.98\% & \tiny 30.16\% & \tiny  29.68\% & \tiny  29.60\%  & \tiny 28.36\% & \tiny  19.80\% & \tiny 11.54\% \\ \cline{2-12}
      & \scriptsize Error & \tiny 5.38\% & \tiny  6.08\% & \tiny  3.79\%  & \tiny 0.42\% & \tiny 3.87\% & \tiny  1.94\% & \tiny  2.87\%  & \tiny 3.72\% & \tiny  12.25\% & \tiny 1.14\% \\ \cline{2-12}
      \multicolumn{9}{r}{}& \multicolumn{2}{|r||}{\tiny \textbf{Average Prediction Error}} & \tiny \textbf{4.15\%} \\ \cline{10-12}
      \multicolumn{12}{r}{} \\ \cline{2-12}
       & \scriptsize Measured & \tiny 48.61\% & \tiny  40.32\% & \tiny  60.80\%  & \tiny 55.04\% & \tiny 47.97\% & \tiny 52.14\% & \tiny 48.38\%  & \tiny 40.58\% & \tiny 23.73\% & \tiny 0.56\% \\ \cline{2-12}
      & \scriptsize Predicted & \tiny 61.78\% & \tiny  61.18\% & \tiny  62.26\%  & \tiny 60.30\% & \tiny 58.66\% & \tiny  57.70\% & \tiny 57.54\%  & \tiny 55.06\% & \tiny 37.94\% & \tiny 21.42\% \\ \cline{2-12}
      & \scriptsize Error & \tiny 13.17\% & \tiny 20.86\% & \tiny 1.46\%  & \tiny 5.26\% & \tiny 10.69\% & \tiny 5.56\% & \tiny 9.16\%  & \tiny 14.48\% & \tiny 14.21\% & \tiny 20.86\% \\ \cline{2-12} 
      \multicolumn{9}{r}{}& \multicolumn{2}{|r||}{\tiny \textbf{Average Prediction Error}} & \tiny \textbf{11.57\%} \\ \cline{10-12}
      \multicolumn{12}{r}{} 
      \end{tabular}\\ \normalsize 
      \caption{Measured and predicted percentages for missing and archived content in each dataset}
      \label{tab:afterayear}
  \end{table}
\end{center}
To verify the second part of our model we calculate the percentages of resources that are archived at least once in one of the public archives.
Table \ref{tab:afterayear} illustrates the archived results measured, predicted, and the corresponding standard error. Figure \ref{fig:afterayear} also displays the measured and predicted 
corresponding plots for the archived resources. While the archived content percentages had a higher error percentage of 11.57\% and 
proceeded to become further less linear with time. This fluctuation in the archival percentages convinced us that a further analysis is needed.

\subsection{Reappearance and Disappearance}
In measuring the percentage of resources missing from the live web, we assumed that when a resource is deemed missing it remains missing. It was also assumed that if a 
snapshot of the resource is present in one of the public archives the resource is deemed archived and that this snapshot persists indefinitely. Utilizing the response logs resulting from running the existence experiment in 2012 and in 2013 we compare 
the corresponding HTTP responses and the number of mementos for each resource. As expected, portions of the datasets disappeared from the live web and were labeled as 
missing. An interesting phenomena occurred as several of the resources that were previously declared as missing became available again as shown in 
table \ref{tab:reappear}. A possible explanation of this reappearance could be a domain or a webserver being disrupted and restored again. Another possible 
explanation is that the previously missing resources could be linked to a suspended user account that was reinstated. To eliminate the effect of transient errors, 
the experiment was repeated three times in the course of two weeks. 
\begin{table}[ht]
    \begin{tabular}{|c||c|c|c|c|c|c|c|}
        \hline
	\textbf{Event} & \textbf{MJ} & \textbf{Iran} & \textbf{Obama } & \textbf{H1N1} & \textbf{ Egypt} & \textbf{Syria} & \textbf{Average} \\ \hline
        \tiny \% Re-appearing on the web \normalsize & 11.29\% & 11.48\% & 6.63\% & 3.68\% & 4.21\% & 1.97 \% &6.54\% \\ \hline
        \tiny \% Disappearing from archives\normalsize & 9.98\% & 11.17\% & 15.65\% & 5.46\% & 2.81\% & 2.25 \% & 7.89\% \\\hline
        \tiny \% Going from 1 memento to 0\normalsize & 2.72\% & 2.89\% & 4.24\% & 1.96\% & 0.23\% & 0.28\% & 2.05\% \\
        \hline
    \end{tabular}\\
        \caption{Percentages of resources reappearing on the live web and disappearing from the public archives per event.}
    \label{tab:reappear}
\end{table}

\begin{figure}
\centering
\includegraphics[scale=0.28]{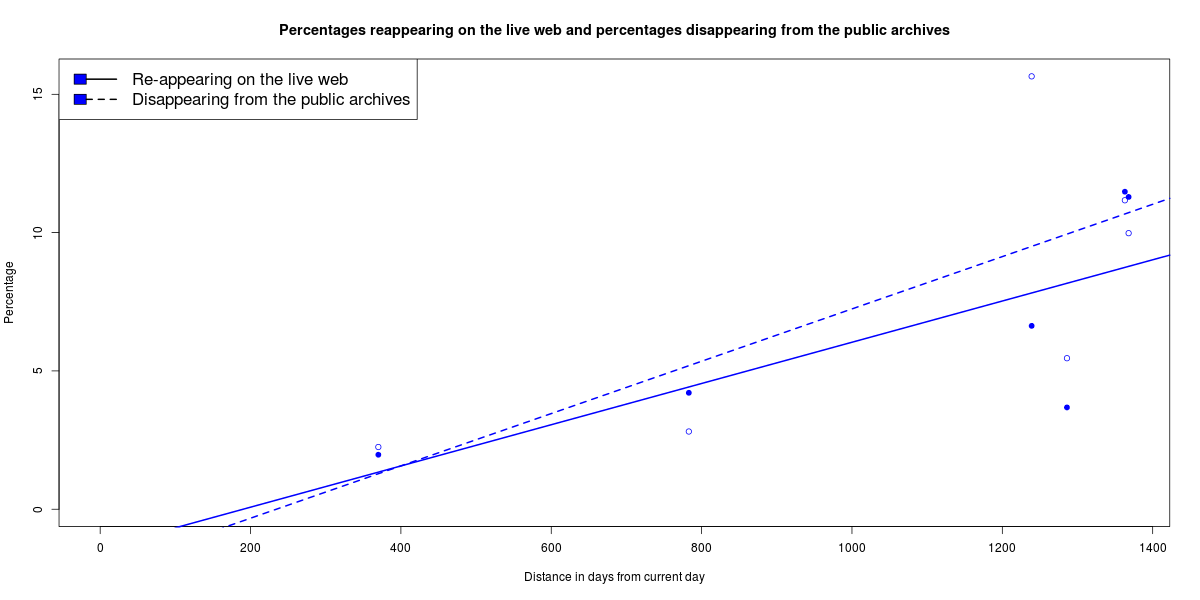}
\caption{Percentages of resources reappearing on the live web and the resources disappearing from the public archives.}
\label{fig:reappeardisappearence}
\end{figure}
The dotted line in figure \ref{fig:reappeardisappearence} shows resources missing in 2012 that reappeared in 2013. Given those percentages we notice a linear relationship with time. By applying linear regression we reach equation \ref{eq:2} describing the reappearance of resources as a function of time.
\begin{equation}
Live Content\hspace{0.8 mm}Reappearing = 0.01 (Age\hspace{0.8 mm}in\hspace{0.8 mm}days) - 1.42
 \label{eq:2}
\end{equation}

In the same previous study, we modeled the archival existence or the percentage archived as a function of time. The phenomena analyzed in the previous section showed the instability of the resources in the 
web which influenced us to investigate the archived resources as well. We deemed a resource to be archived if there existed at least one publicly available memento of the resource in the archives. For each 
resource we extracted the memento timemaps and recorded the number of available mementos. The resources are expected to have the same number of mementos or more indicating more snapshots taken into the 
archives or unarchived resources started to exist in the archives. We notice another interesting phenomena, the number of available mementos of several resources have actually 
descreased indicating disappearance from the archives as shown in table \ref{tab:reappear}. 

Brunelle and Nelson have shown that timemaps shrink 20\% of the time \cite{jcdl2013justin}. Another possible explanation is that in 2012 the memento aggregator included search engine caches as archives but no longer does so in 2013. We estimate search engine cache only timemaps by measuring the number of resources whose timemaps went from 1 memento to 0 as shown in table \ref{tab:reappear} as well. 
Similarly, we plot the percentages 
of archival disappearance in figure \ref{fig:reappeardisappearence}. Equation \ref{eq:3} results from applying linear regression in curve fitting. Inspecting figure \ref{fig:reappeardisappearence} verifies to a certain degree our 
explanation of the archival disappearance phenomena as the regression line maintains the same slope of the estimated model as shown in figure \ref{fig:afterayear} while it differs in the Y-intercept. This explains to a certain degree the uniform variation in the estimated function. Unfortunately, we cannot verify this precisely as we do not have the past timemaps of the resources in the datasets. 

\begin{equation}
Mementos\hspace{0.8 mm}Disappearing = 0.01 (Age\hspace{0.8 mm}in\hspace{0.8 mm}days) - 2.22
 \label{eq:3}
\end{equation}
\subsection{Tweet Existence}
\begin{figure}
\centering
\includegraphics[scale=0.30]{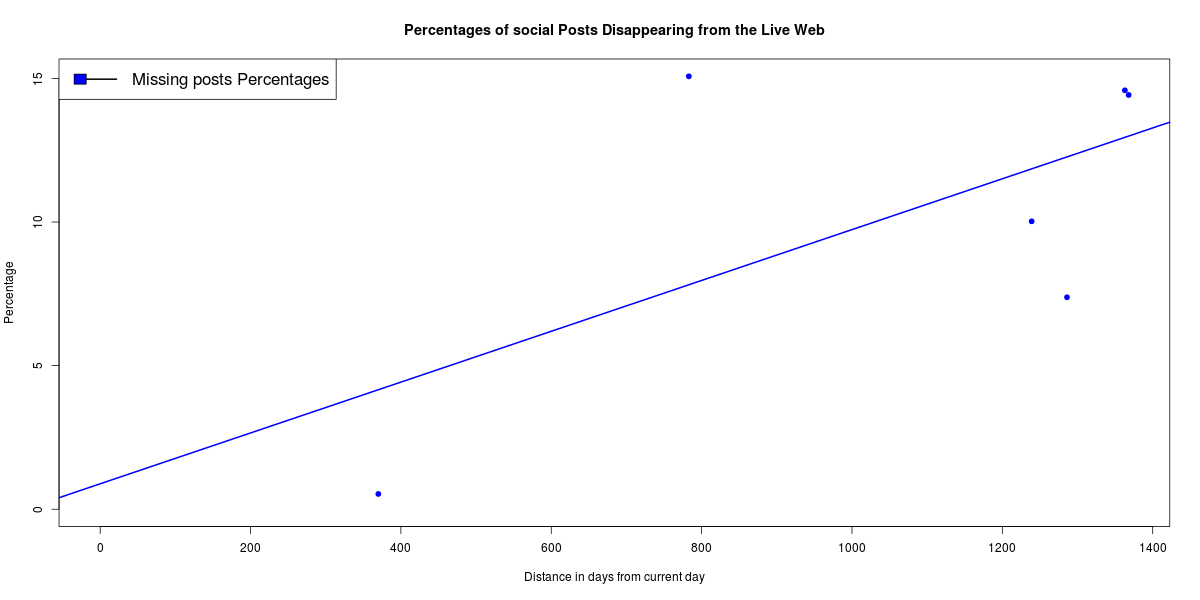}
\caption{Percentages of missing posts averages curve fitted using linear regression.}
\label{fig:posts}
\end{figure}

After focusing on the embedded resources shared in posts in social media another question arouse, what about the existence of the social post itself? In collecting the dataset that we utilized in our analysis 
we focused on the embedded resource and the creation dates. Also the Stanford Network Analysis Platform (SNAP) dataset we used provides only the tweet text, the author's username, and the creation date with no further information about the tweet or its URI. A 
social post could face the same fate of the embedded resource by being deleted, service hosting it discontinued, or the author's account getting suspended. Similarly to the resource existence testing, we check the existence of the posts by 
examining the HTTP response headers. Unfortunately, the datasets we used do not include all the fields and parameters of a tweet, among which is the tweet's URI. To work around the absence of the social post URI we utilized Topsy, a service that mines social media websites like Twitter to provide analytics and insight to topics and resources. Using 
the API, we can extract all the available tweets that incorporate a given URI with a maximum of 500 tweets. For each resource in the dataset we extract all the tweets and check their existence on the live web accordingly. Given 
a URI, we can estimate the percentage of social posts that are missing. This number could give an insight to what is the probability that the post itself is missing. Table \ref{tab:3} shows the results for 
each dataset. Figure \ref{fig:posts} illustrates the collective percentages through time. Equation \ref{eq:4} shows the result of curve fitting the percentages of loss as a function of time.
\begin{table}[ht]
    \begin{tabular}{|c||c|c|c|c|c|c|c|}
        \hline
	\textbf{Event} & \textbf{MJ} & \textbf{Iran} & \textbf{Obama } & \textbf{H1N1} & \textbf{ Egypt} & \textbf{Syria} & \textbf{Average} \\ \hline
        Average \% of missing posts & 14.43\% & 14.59\% & 10.03\% & 7.38\% & 15.08\% & 0.53\% & 10.34\% \\
        \hline
    \end{tabular}
        \caption{Percentages of missing posts averages.}
\label{tab:3}
\end{table}
\begin{equation}
Social Posts\hspace{0.8 mm}Missing = 0.01 (Age\hspace{0.8 mm}in\hspace{0.8 mm}days) + 0.88
 \label{eq:4}
\end{equation}

\section{Context Discovery and Shared Resource Replacement}
A web resource can fall into one of the categories as shown in table \ref{tab:2}. These categories were adopted from the work of McCown and Nelson \cite{frankframework}.
\begin{table}[ht]
  \centering
    \begin{tabular}{c|c|c|}
    \cline{2-3}
      & \textbf{Archived} & \textbf{Not Archived} \\ \cline{2-3}
    \textbf{Available} & Replicated& Vulnerable\\ \hline
    \textbf{Missing} & Endangered& Unrecoverable\\ \cline{2-3}
    \end{tabular}
    \label{tab:2}
    \caption{Web resource categories in regards to archivability and availability}
\end{table}

If a resource is available on the live web and also archived in public archives then it is considered replicated and safe. 
The resource is considered vulnerable if it persists on the web but has no available archived versions. If a resource is not available on the live web but has 
an archived version then it is considered endangered as it relies on the stability and the persistence of the archive. The worst case scenario occurs when 
the resource disappears from the live web without being archived at all thusly, be considered unrecoverable. In our study we focus on the latter category and how we can utilize the social media in identifying the context 
of the shared resource and select a possible replacement candidate to fill in the position of the missing resource and maintain the same context of the social post.

A shared resource leaves traces even after it ceases to exist on the web. We attempt to collect those traces and discover context for the missing resource. Since Twitter for example restricts the length of the posts to 
be 140 characters only, an author might rely mostly on the shared resource in conveying a thought or an idea by embedding a link in the post and resorting to limiting the associated text. Thusly, obtaining context is crucial when the resource disappears.
To accomplish that, we try to find the social link neighborhood of the tweet and the resource we are attempting this context discovery. When a link is shared on Twitter for example, it could be associated with describing text in the form 
of the status itself, hashtags, usertags, or other links as well. These co-existing links could act as a viable replacement to the missing resource under investigation while the tags and text could provide better context enabling a better 
understanding of the resource.

\subsection{Social Extraction}
Given the URI of the resource under investigation, we utilize Topsy's API to extract all the available tweets incorporating this URI. In social media, a resource's URI can be shared in different forms with the aid of URL shortening. 
To elaborate, a link to Google's web page http://www.google.com could be shared also in several forms like http://goo.gl/xYMol, http://bitly.com/XeRH58, and http://t.co/XFiAkbHnp3. Each of these forms redirects to the same final destination URI. Fortunately, Topsy's API handles 
this by searching their index for the final target URL rather than the shortened form. A maximum of 500 tweets of the most recent tweets posted could be extracted from the API regarding a certain URL. The content from all the tweets is collected to 
form a social context corpus.

From this corpus we extract the best replacement tweet by calculating the longest common N-gram. This represents the tweet with the most information that describes the target resource intended by the author. Within some tweets, multiple links 
coexist within the same text. These co-occurring resources share the same context and maintain a certain relevancy in most cases. A list of those co-occurring resources are extracted and filtered for redundancies. Finally, the textual components of the tweets 
are extracted after removing usertags, URIs, social interaction symbols like ``RT''. We named the document composed of those text-only tweets in the form of phrases the \textit{``Tweet Document''}. 

Figure \ref{fig:json} illustrates the JSON object produced from social mining the resource as described above.
\begin{figure}[ht]
\centering
\begin{Verbatim}[]
Reconstruction:
{
  "URI": "http://ws-dl.blogspot.com/2012/02/2012-02-11-losing-my-revolution
  -year.html",
  "Related Tweet Count": 290,
  "Related Hashtags": "#history #jan25 #sschat #arabspring #jrn112 #archives 
  #in #revolution
  #iipc12 #mppdigital #egypt #recordkeeping #twitter #egyptrevolution 
  #digitalpreservation 
  #preservation #webarchiving #or2012 #1anpa #socialmedia",
  "Users who talked about this": "@textfiles @jigarmehta @blakehounshell] 
  @jonathanglick 
  @daensen404: @ryersonjourn @chanders @theotypes) @jwax55 @marklittlenews 
  @ndiipp ...",
  "All associated unique links:": "http://t.co/ZRASTg5o http://t.co/eXhlSTRF   
  http://t.co/3GIb6oI3 http://t.co/ArVqCqfP ...",
  "All other links associated:": "http://www.cs.odu.edu/~mln/pubs/tpdl-
  2012/tpdl-2012.pdf 
  http://dashes.com/anil/2011/01/if-you-didnt-blog-it-it-didnt-happen.html",
  "Most frequent link appearing:": "http://t.co/0A1q2fzz",
  "Number of times the Most frequent link appearing:": 19,
  "Most frequent tweet posted and reposted:": "@acarvin You may have 
  seen this already. 
  Arab Spring digital content is apparently being lost.",
  "Number of times the Most frequent tweet appearing:": 23,
  "The longest common phrase appearing:": "You may have seen this already 
  Arab Spring 
  digital content is apparently being lost",
  "Number of times the Most common phrase appearing:": 28
}
\end{Verbatim}
\caption{Social Content Extraction using Topsy API}
\label{fig:json}
\end{figure}

\subsection{Resource Replacement Recommendation}
From the social extraction phase above we gathered information that helps us to infer the aboutness and context of a resource. Given this context, can we utilize it in obtaining a viable replacement resource to 
fill in the missing one and provide the same context? 

To answer this, we utilize the work of Klein and Nelson \cite{lexical} in defining the lexical signatures of web pages as discussed earlier. First, we extract the tweet document as described above. Next, we remove all the stop words 
and apply Porter's stemmer to all the remaining words \cite{porter}. We calculate the term frequency of each stemmed word and sort them from highest occurring to the lowest. Finally, 
we extract the top five words to form our tweet signature.

On the one hand, and using this tweet signature as a query, we utilize Google's search engine to extract the top 10 resulting resources. On the other hand, we collect all the other co-occurring pages in the tweets obtained by the API. 
These pages combined produce a replacement candidate list of resources. One or more of which can be utilized as a viable replacement of the resource under investigation.

To choose which resource is more relevant and a possibly better replacement we utilize once more the tweet document extracted earlier. For each of the extracted pages in the candidate list, we download the representation and utilize the boilerpipe 
library in extracting the text within\footnote{http://code.google.com/p/boilerpipe/}. The library provides algorithms to detect and remove the ``clutter'' (boilerplate, templates) around the main textual content of a web page. 
Having a list of possible candidate textual documents and the tweet document, the next step is to calculate similarity. The pages are sorted according to the cosine similarity to the tweets page 
describing the resource under reconstruction. 

At this stage we have extracted contextual information about the resource and a possible replacement. The next step is to measure how well the reconstruction process was undergone and 
how close is this replacement page is to the missing resource.
\section{Evaluation}
Since we cannot measure the quality of the discovered context or the resulting replacement page to the missing resource, we have to set some assumptions. We extract a dataset of resources that are currently available on the live web and 
assume they no longer exist. Each 
of these resources are textual based and neither media files nor executables. Each of these resources has to have at least 30 retrievable tweets using Topsy's API to be enough to build context.

\begin{figure}
\centering
\includegraphics[scale=0.3]{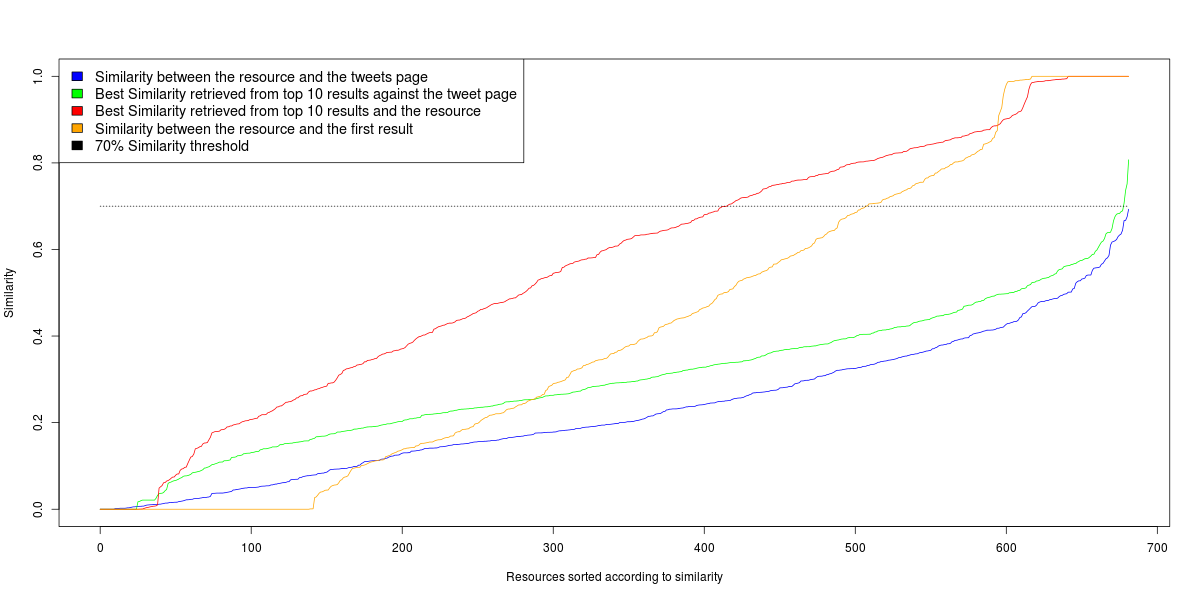}
\caption{Similarities with the original resource $R_{missing}$}
\label{fig:eval}
\end{figure}

We collect a dataset of 731 unique resources following these rules. We perform the context extraction and the replacement recommendation phases. We download the resource under investigation ($R_{missing}$) and the list of candidate replacements from the search engines 
($R_{search}$) and the list of co-occurring resources ($R_{co-occurring}$). For each we use the boilerpipe library to extract text and use cosine similarity to perform the comparisons. For each resource, we measure the similarity between the ($R_{missing}$) and the 
extracted tweet page. For each element in ($R_{search}$) we calculate the cosine similarity with the tweet page and sort the results accordingly from most similar to the least. We repeat the same with the list of co-occurring resources ($R_{co-occurring}$). 
Then we calculate the similarity between ($R_{missing}$) and ($R_{search}(first)$) indicating the top result obtained from the search engine index. Then, we compare ($R_{missing}$) with each of the elements in ($R_{search}$) and ($R_{co-occurring}$) to demonstrate the best possible similarity. Figure \ref{fig:eval} illustrates the different 
similarities sorted for each measure and shows that 41\% of the time we can extract a significantly similar replacement page ($R_{replacement}$) to the original resource ($R_{missing}$) by at least 70\% similarity. Finally, we needed to validate the 
effectiveness of using the tweet signature as a query string to the search engine. Using the tweet signature extracted from tweets associated with an existing resource against the search engine API and locating the rank in which the resource appear in the results list, we calculate the mean reciprocal rank to be 0.43.
\section{Conclusions and Future Work}
In this study we verify our previous analysis and estimation of the percentage missing of the resources shared on social media. The function in time still holds in modeling the percentage disappearing from the web. As for the model estimated for the amount 
archived it showed an alteration. The slope of the regression line in the model stayed the same while the y-intercept varied. We deduce that a possible explanation to this phenomena is due to timemap shrinkage. Previously, timemaps incorporated search engine 
caches as mementos which was removed in the most recent Memento revision. Next, we classified web resources into four different categories in regards to existence on the live web and in public web archives. Then we considered the unrecoverable category where the resource is deemed missing from the live web whilst not having any 
archived versions. Since we cannot perform a full reconstruction or retrieval, we utilize the social nature of the shared resources by using Topsy's API in discovering the resource's context. Using this context and the co-occurring resources we apply a 
range of heuristics and comparisons to extract the most viable replacement to the missing resource from its social neighborhood. Finally, we performed an evaluation to measure the quality of this replacement and found that for 41\% of the 
resources we can obtain a significantly similar replacement resource with at least 70\% similarity.
For our future work, we would like to expand our investigation to incorporate other resources of different types like images and videos. A further investigation is crucial to better rank the results and account for the different types of resources.
\section{Acknowledgments}
This work was supported in part by the Library of Congress and NSF IIS-1009392.

\end{document}